\documentclass{elsart}
\usepackage{epsfig}
\usepackage{rotate}

\begin{document}
\newcommand\tanlam{\tan\lambda}


\newcommand\vm{m}
\newcommand\veps{\varepsilon}

\newcommand\vbeta{\mbox{\boldmath$\beta$}}
\newcommand\valpha{\mbox{\boldmath$\alpha$}}
\newcommand\valphaone{\mbox{\boldmath$\alpha$}_\mathrm{I}}
\newcommand\valphatwo{\mbox{\boldmath$\alpha$}_\mathrm{II}}
\newcommand\valphathree{\mbox{\boldmath$\alpha$}_\mathrm{III}}
\newcommand\vB{\mathbf{B}}
\newcommand\vw{\mathbf{w}}
\newcommand\vf{\mathbf{f}}

\newcommand\vr{\mathbf{r}}
\newcommand\vx{\mathbf{x}}
\newcommand\sx{\mathbf{\widehat{x}}}
\newcommand\vxmin{\sx_k}
\newcommand\vxf{\vx^F}
\newcommand\vxfmin{\vxmin^{F}}
\newcommand\vxb{\vx^B}
\newcommand\vxbmin{\vxmin^{B}}

\newcommand\chisq{\chi^2}
\newcommand\chif{\chi^{2~(F)}}
\newcommand\chifmin{\chi^{2~(F)}_k}
\newcommand\chib{\chi^{2~(B)}}
\newcommand\chibmin{\chi^{2~(B)}_k}

\newcommand\chikfb{\chi^{2~(FB)}_k}
\newcommand\maxchikfb{\widetilde\chi^{2~(FB)}}

\newcommand\fullsix{\mathrm{full,0}}
\newcommand\chifullsix{\chi^2_{\rm trk}} 
\newcommand\chifullsixmin{\chi^2}
\newcommand\fullseven{\mathrm{full,I}}
\newcommand\chifullsevenk{\chi^2_{{\fullseven},k}}
\newcommand\minchifullseven{\widetilde\chi^2_{\fullseven}}
\newcommand\fulleight{\mathrm{full,II}}
\newcommand\chifulleightk{\chi^2_{{\fulleight},k}}
\newcommand\minchifulleight{\widetilde\chi^2_{\fulleight}}
\newcommand\fullnine{\mathrm{full,III}}
\newcommand\chifullninek{\chi^2_{{\fullnine},k}}
\newcommand\minchifullnine{\widetilde\chi^2_{\fullnine}}

\newcommand\DIR{D_{\mathrm{I},k}(1/R)}
\newcommand\DIIphi{D_{\mathrm{II},k}(\phi)}
\newcommand\DIIlam{D_{\mathrm{II},k}(\tanlam)}
\newcommand\DIIIR{D_{\mathrm{III},k}(1/R)}
\newcommand\DIIIlam{D_{\mathrm{III},k}(\tanlam)}
\newcommand\DIIIphi{D_{\mathrm{III},k}(\phi)}

\newcommand\FI{F_{\mathrm{I},k}}
\newcommand\FII{F_{\mathrm{II},k}}
\newcommand\FIII{F_{\mathrm{III},k}}

\newcommand\minF{\widetilde F}
\newcommand\minFI{\widetilde F_{\mathrm{I}}}
\newcommand\minFII{\widetilde F_{\mathrm{II}}}
\newcommand\minFIII{\widetilde F_{\mathrm{III}}}

\newcommand\dof{\mathrm{dof}}

\newcommand\ddata[1]{m_{#1}}
\newcommand\dtheo[1]{h_{#1}(\vx_{#1})}
\newcommand\vddata{\mathbf{m}}
\newcommand\vdtheo{\mathbf{h}}
\newcommand\vddataF{\vddata^{F}}
\newcommand\vdtheoF{\vdtheo^{F}}
\newcommand\vddataB{\vddata^{B}}
\newcommand\vdtheoB{\vdtheo^{B}}

\newcommand\Vd{V^{(m)}}
\newcommand\VdF{V^{(m,F)}}
\newcommand\VdB{V^{(m,B)}}

\newcommand\Vx{V^{(\sx)}}
\newcommand\VxF{V^{(\sx_k,F)}}
\newcommand\VxB{V^{(\sx_k,B)}}

\newcommand\xfp{x^{\mathrm{\,pred},F}_i(\valpha)}
\newcommand\xbp{x^{\mathrm{\,pred},B}_i(\valpha)}
\newcommand\vxfp{\vx^{\mathrm{\,pred},F}(\valpha)}
\newcommand\vxbp{\vx^{\mathrm{\,pred},B}(\valpha)}

\newcommand\numucc{\numu \cc}
\newcommand\numunc{\numu {\rm NC}}
\newcommand\pimunu{\pi\rightarrow\mu\nu}

\newcommand\epsdir{/usr/home/user1/tanya/elsevier}

\begin{frontmatter}

\title{Kalman Filter Track Fits and \\Track Breakpoint Analysis}
\author[Paris]{Pierre Astier}
\author[UCLA]{Alessandro Cardini}
\author[UCLA]{Robert D. Cousins\thanksref{email}}
\author[Paris]{Antoine Letessier-Selvon}
\author[Paris]{Boris A. Popov\thanksref{Dubna}}
\author[UCLA]{Tatiana Vinogradova}

\address[Paris]{LPNHE, Laboratoire de Physique Nucl\'eaire et 
des Hautes Energies,\\
Universit\'es de Paris 6 et 7, 75252 Paris Cedex 05, France.}
\address[UCLA]{Department of Physics and Astronomy,
University of California,\\
Los Angeles, California 90095, U.S.A.}
\thanks[email]{Email address: cousins@physics.ucla.edu}
\thanks[Dubna]{on leave from the Laboratory of Nuclear Problems, JINR, 
141980 Dubna, Russia.}

\begin{abstract}
We give an overview of track fitting using the Kalman filter method in
the NOMAD detector at CERN, and emphasize how the wealth of by-product
information can be used to analyze track breakpoints (discontinuities
in track parameters caused by scattering, decay, etc.).  After
reviewing how this information has been previously exploited by
others, we describe extensions which add power to breakpoint detection
and characterization.  We show how complete fits to the entire track,
with breakpoint parameters added, can be easily obtained from the
information from unbroken fits.  Tests inspired by the Fisher $F$-test
can then be used to judge breakpoints.  Signed quantities (such as
change in momentum at the breakpoint) can supplement unsigned
quantities such as the various chisquares.  We illustrate the method
with electrons from real data, and with Monte Carlo simulations of
pion decays.
\end{abstract}
\begin{keyword} 
breakpoints, Kalman filter, track fitting\\
PACS code: 07.05.Kf 
\end{keyword}
\end{frontmatter}


\section{Introduction}
\label{sec:intro}

The Kalman filter is an efficient algorithm for fitting tracks in
particle spectrometers with many position-sensing detectors
\cite{Billoir84,Billoir85,Fruh,BilloirQian,Stampfer,Regler}.  It cures
many of the problems of traditional $\chisq$ track fitting using
Newton steps, which becomes more and more unwieldy as the number of
position measurements increases. In such situations where Kalman
filtering is naturally applied, it can be possible to detect and
characterize track {\em breakpoints}, defined as locations where one
or more of the track parameters is discontinuous.  Obvious
breakpoints, such as a large kink due to a particle decay, are often
found before a full track fit is performed.  More subtle breakpoints
may only manifest themselves when the track is fit to obtain the track
parameters.

Fr\"uhwirth\cite{Fruh} has investigated the detection of breakpoints
using information which is a natural by-product of a Kalman filter
track fit.  When fitting a drift-chamber track with $N$ position
measurements (``hits''), the idea has been the following.  At the
location of hit $k$, one has the best-fit track parameters (a) using
hits 1 through $k$ and (b) using hits $k+1$ through $N$.  One
constructs a $\chisq$ for the consistency of these two sets of track
parameters.  This $\chisq$, along with other $\chisq$'s at hand from
the two fits, can be combined to form test statistics for breakpoints.

Such methods suffer a loss of power because of two defects.  First,
$\chisq$'s by construction throw away information about the arithmetic
signs of differences; such information is relevant since a track's
momentum should normally {\em decrease} when the particle decays.

Second, appropriate constraints of a breakpoint hypothesis are not
incorporated.  For example, if a particle decays at hit $k$, a desired
quantity is the mismatch in track parameters describing the {\em
momentum} vector, {\em under the constraint} that the track {\em
position} vector from fits (a) and (b) is identical.

In this paper, we describe a procedure which uses information from the
Kalman filter fit to construct the result of a {\em full} track fit
which has additional track parameters to account for the
discontinuities at a breakpoint.  It is natural to allow for one, two,
or three discontinuous parameters in order to describe different
physical processes:
\begin{description}
\item{\bf Type I:} An electron emitting a bremsstrahlung photon 
generally
changes only its momentum {\em magnitude},
since the photon is essentially collinear with
the electron direction.
\item{\bf Type II:} A particle with a hard elastic scatter may have 
momentum magnitude essentially unchanged, while changing the two
angles specifying direction.
\item{\bf Type III:} A charged pion or kaon decaying to $\mu\nu$ in 
general changes momentum magnitude as well as the two angles.
\end{description}
With the method we describe, trial breakpoint fits at every hit (away
from the ends) of the track can be quickly obtained, and used to
search for and characterize breakpoints.

Billoir \cite{Billoir84} has investigated Type II breakpoints,
performing fits which do not assume an existing breakpointless fit.
We show in this paper how the by-products of a Kalman filter fit allow
one to avoid refitting the hits while incorporating the constraints.

We discuss these tools in the context of the NOMAD \cite{NOMAD}
neutrino experiment at CERN, within which this development took place.
However, the results are generally applicable to any experiment in
which the number of position measurements is large enough to fit the
track on both sides of the potential breakpoint.

In Sec.~\ref{sec:param} we introduce some notation and describe our
track parameters and track models, including energy loss.  In
Sec.~\ref{sec:fit} a review of the traditional (non-Kalman) track fit
is given. Sec.~\ref{sec:kalman} describes its replacement by the
Kalman filter.  In Sec.~\ref{sec:earlier}, we briefly review previous
work on breakpoint variables.  In Sec.~\ref{sec:additional}, we
introduce the new breakpoint variables, and in Sec.~\ref{sec:results},
we present some indicative results of their use.  We conclude in
Sec.~\ref{sec:conclusion}.

\section{Track parameters and track models}
\label{sec:param}
\subsection{Parameters}
When fitting a track, one typically describes its location in 6D phase
space by choosing a fixed reference surface (a vacuum window, chamber
plane, etc.)  and then fitting for the 5 independent parameters of the
track at the position {\em where the track intersects this surface}.
We let the vector $\vx$ contain these track parameters.  In a
fixed-target experiment with beam direction along the $z$ axis, the
parameterization of $\vx$ is often taken to be
\begin{equation} \vx=(x,y,\d x/\d z,\d y/\d z,q/p) \end{equation}
at a reference plane at fixed $z$, where $q/p$ is the charge/momentum.

\begin{figure}
\centering\epsfig{file=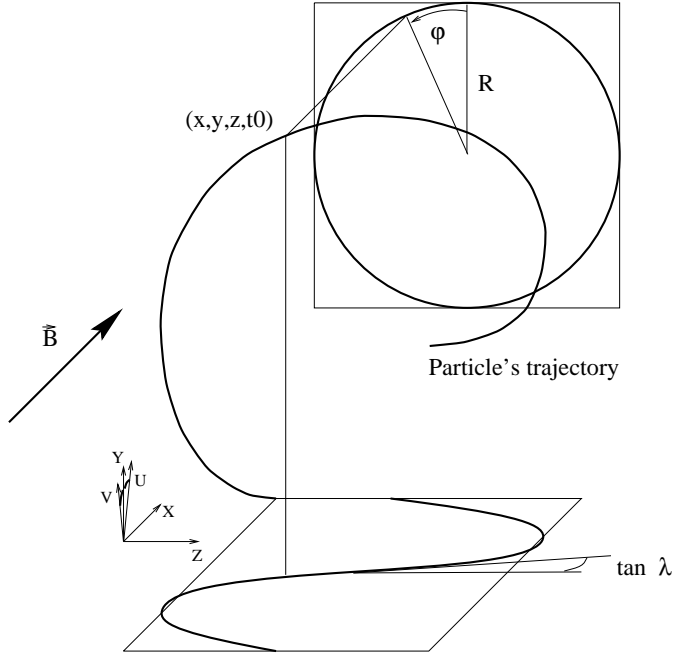,width=90mm}
\caption{Definition of the helix parameters used to describe charged 
particle trajectory in the NOMAD setup.}
\label{fig:helix_params}
\end{figure}

NOMAD is a fixed-target experiment with drift chamber planes
perpendicular to the $z$ direction (nearly aligned with the neutrino
beam).  However, the chambers are immersed in a uniform magnetic
field\footnote{Sense wires of one chamber make angles of +5, 0 and --5
degrees with respect to the magnetic field direction providing a space
measurement along coordinates designated $u$, $y$, and $v$.}, so that
soft tracks often loop back, and a helical parameterization similar to
collider experiments is more appropriate. We maintain the reference
surface as a plane with fixed $z$ (the $z$ of the first hit of the
track), and specify a track there by
\begin{equation}
\label{eqn:xparams}
 \vx=(x,y,1/R,\tanlam,\phi, t),
\end{equation}
where we have introduced the three parameters of a helical curve in
the uniform magnetic field: the signed\footnote{In NOMAD, $1/R$ has a
sign opposite to the particle charge.}  inverse radius of curvature
$1/R$, the dip angle $\tanlam$, and the angle of rotation $\phi$~(see
Fig.~\ref{fig:helix_params}). In addition to these 5 traditional
parameters, NOMAD has a sixth parameter, the zero-time-offset for the
drift chamber measurements, called $t$.  It has been introduced
because of trigger time jitters. In the following, unless specified
otherwise, {\it track parameters} will refer to this second
parameterization.

The parameter $1/R$ is related to the momentum $p$ by
\begin{equation}
\frac{1}{R} \propto \frac{B}{p}\sqrt{1 + (\tan\lambda)^2},
\end{equation}
while the sign of ($1/R$) only reflects the particle charge if the
time runs the right way along the track. In NOMAD the sign convention
we used implied that the product $R \cdot \phi$ increases with time
along the assumed time direction given by the ordering of the
measurements along the track.

Throughout our work, we use a change in $1/R$ as an indicator of a
change in $p$; this is strictly true only when the change in
$\tan\lambda$ is negligible, but is an adequate approximation.
(Inhomogeneity in the magnetic field B does not matter, since we
compare $1/R$ estimates at the same point.)

With this parameterization, the three physical processes enumerated in
the Introduction have the following breakpoint signatures:
\begin{description}
\item{\bf Type I:} An electron emitting a bremsstrahlung photon
has a discontinuity in $1/R$.
\item{\bf Type II:} A pion with a hard elastic scatter has 
discontinuities in
$\tan\lambda$ and $\phi$.
\item{\bf Type III:} A charged pion or kaon decaying to $\mu\nu$ 
has discontinuities in $1/R$, $\tan\lambda$, and $\phi$.
\end{description}

From the estimation of the track parameters at the reference plane one
needs a transformation, called the track model, with which one
computes the expected measurements at any position in the detector.
This model describes the dependence of the measurements on the initial
values in the ideal case of no measurement errors and of deterministic
interactions of the particle with matter.  Use of a correct
description is of the utmost importance for the performance of the
fitting procedure, be it traditional or not.

\subsection{Equation of motion in a magnetic field}
The trajectory of a charged particle in a (static) magnetic field is
determined by the following equation of motion:
\begin{equation}
 \d^2 \vr /\d^2 s = (k q/p) \cdot (\d \vr /\d s) 
\times \vB(\vr (s)), 
\end{equation}
where $\vr$ is the position vector, $s$ is the path length, $k$ is a
constant of proportionality, $q$ is the (signed) charge of the
particle, $p$ is the absolute value of its momentum, and $\vB(\vr)$ is
the static magnetic field.

With our parameterization it proved convenient to use $\phi$ as the
running parameter rather than $z$ since particles may loop back in the
detector and cross the same measurement plane several times.  The
equation of motion can be readily integrated along a trajectory step
(from position 0 to position 1) where $R$ and $\tan \lambda$ are
assumed constant.
\begin{eqnarray}
&x_1  	 &=  x_0 +  R_0 \cdot \tan \lambda \cdot (\phi_1 - \phi_0)\\
&y_1    &=  y_0 +  R_0 \cdot (\cos \phi_1 - \cos \phi_0) \\
&z_1    &=  z_0 +  R_0 \cdot (\sin \phi_1 - \sin \phi_0)
\label{eqn-sinphi}\\
&R_1    	&=  R_0 \\
&\tan\lambda_1 &= \tan \lambda_0 \\
&t_1          &=  t_0 +  R_0\cdot(\phi_1-\phi_0)/(\beta\cos\lambda_0).
\end{eqnarray}
In the last equation, $\beta$ is the particle velocity and $R \cdot
\Delta\phi$ has the right sign following our convention.

In NOMAD, detector measurement planes are located at fixed $z$ and
Eqn.~\ref{eqn-sinphi} can be solved,
\begin{equation}
 \sin\phi_1 = \sin\phi_0 + (z_1 - z_0) \cdot (1/R_0),  
\end{equation} 
to obtain $\phi_1$ at the desired $z$.  In practice, among all the
possible solutions, our track model returns the one corresponding to
the next crossing in the requested time direction.

The magnetic field strength varies by a few percent in the tracking
volume.  This was accommodated by ignoring the minor components of the
field and updating $1/R$ at every tracking step so that the product $R
\cdot B$ remains constant, up to energy losses which are now
discussed.

\subsection{Energy losses}

The ionization losses are accounted for by updating $1/R$ at every
tracking step\footnote{The absolute value derives from our sign
conventions, so that a track gains energy if tracked backwards in
time.}:
\begin{equation}
 \Delta (1/R) = \frac{\d(1/R)}{\d E} \frac{\d E}{\d x}\Delta x =
\frac{|1/R|}{0.3 B \beta} \frac{\d E}{\d x} \Delta \phi
\end{equation}
where $\Delta \phi = \phi_{end} - \phi_{start}$, and where $\d E/\d x$
is given by the Bethe-Bloch equation, evaluated (by default) with the
pion mass, and a local matter density extracted from the detector
model used in the GEANT \cite{GEANT} simulation of the experiment. In
the central tracker part of the detector, the matter density is about
0.1 g/cm$^3$ : the ionization loss model does not need to include
detailed relativistic corrections.

In NOMAD, bremsstrahlung losses should be accounted for (on average)
in the track model for electron tracks, because the central tracker
amounts to about 1 radiation length ($X_0$). In the electron (and
positron) track model, one adds to the ionization losses (evaluated
with the electron mass) bremsstrahlung losses:

\begin{equation}
 \Delta (1/R) = \frac{\Delta \phi}{\beta^2 X_0 \cos{\lambda}},
\end{equation}
where we can readily approximate $\beta = 1$. As expressed, these
losses include the whole radiated photon spectrum, although beyond a
certain threshold, the radiated photons can be detected in the
downstream electromagnetic calorimeter, or sometimes as a conversion
pair in the tracker.  But no convincing way was found to define a
threshold that separates continuous small losses from accidental big
ones.

\section{Traditional track fits}
\label{sec:fit}
Given the track model and an estimate ($\vx_0$) of the parameters at
the reference position, one can compute at each measurement location
in the detector (labeled $k = 1,\ldots N$) the theoretical ``{\em
ideal measurements}'' which would be made in the absence of
fluctuations due to the two categories of ``noise'': process noise
(multiple scattering, bremsstrahlung, etc.) and measurement noise
(detector resolution).  This computation is represented by the system
equation in the absence of noise,
\begin{equation}
 \vx_k = \vf_k(\vx_0),
\label{eqn:sys}
\end{equation}
where $\vf$ is a deterministic function (the track model) giving the
values of the parameters at each location $k$.

In general the set of parameters $\vx_k$ is not measured directly by
the apparatus; only a function of it, $h_k(\vx_k)$, is observed.  (In
our case $h_k$ is a drift chamber position measurement.)  Let
\begin{equation} 
\vm_k \ = \ h_k(\vx_k) \ + \ \veps_k
\label{eqn:meas}
\end{equation}
be the measurement equation where $\veps_k$ represents the measurement
errors. By convention, $\langle \vm_k \rangle \ = \ h_k(\vx_k)$ and
$\langle \veps_i\rangle = 0 $. In practice, one has to check that the
track model (involved in the calculation of $\vx_k$) and the
measurement function $h_k$ fulfill this convention.

In a traditional (non-Kalman) track fit, one calculates the
$N${}$\times${}$N$ covariance matrix of the measurements\footnote{The
superscript in parentheses in $\Vd$ is to make clear which vector it
corresponds to; this will be the convention in this paper.}:

\begin{equation}
\Vd_{ij} = \langle (\vm_i - h_i(\vx_i))(\vm_j - h_j(\vx_j))\rangle,
\label{eqn:Vd}
\end{equation}
where the angle-brackets represent an average over an ensemble of
tracks with the same track parameters $\vx$.  According to the
Gauss-Markov theorem, the minimization of
\begin{equation}
\label{eqn:chifull}
\chifullsix(\vx) = \left (\vddata - \vdtheo(\vx)\right )^T\ [\Vd]^{-1}\
           \left (\vddata - \vdtheo(\vx)\right )
\end{equation}
yields parameter estimates with minimum variance among all linear
unbiased estimates.

Note that when the parameter evolution is not affected by any
stochastic noise (as assumed by Eq.~\ref{eqn:sys}), 
$\Vd_{ij} = \langle \veps_i\ \veps_j \rangle$ 
is diagonal (at least block diagonal, if the apparatus provides
multidimensional correlated measurements), and may depend weakly on
$\vx$ via the detector response (for example, the spatial resolution
of drift chambers depends on the track angle w.r.t the anode plane).

The system equation becomes non deterministic when the track
experiences stochastic processes such as multiple scattering,
bremsstrahlung or ionization losses. The system equation
(Eq.~\ref{eqn:sys}) becomes :
\begin{equation}
 \vx_k = \vf_k(\vx_0) +\vw_k
\label{eqn:sysb}
\end{equation}
where $\vw_k$ is a random vector representing the fluctuation of the
parameters along the path from location 0 to location $k$. This {\em
process noise} translates, via the track model, into off-diagonal
elements in $\Vd$ (the noise from location $0$ to $k$ depends on the
noise from location $0$ to $k-1$ and $k-1$ to $k$).

More importantly, $\Vd$ now depends on the reference location $z_0$ at
which the parameters will be estimated.  Of course the average value
of the perturbing effects have to be included in $\vf_k$ but the
fluctuations (e.g. track scattering or energy loss straggling when
relevant) have to be described by the p.d.f of $\vw_k$.  In fact only
the covariance matrix of the $\vw_k$ is needed in practice.  (See
Sec.~\ref{sec:kalman}.)

Given particular data $\vddata$, the track fit consists in finding the
value of $\vx$ which minimizes Eq.~\ref{eqn:chifull}.  Minimization is
typically an iterative process with some convergence criteria to
decide when to stop iterating.  The matrix $\Vd$ can be calculated by
Monte Carlo techniques or sometimes analytically.  If one is
fortunate, the calculation depends only weakly on $\vx$, so it can be
done once per fit using the initial guess for $\vx$, and not changed
at each iteration.  Note that the $N${}$\times${}$N$ matrix $\Vd$ must
be inverted, where $N$ is the number of measurement positions.

Let $\chifullsix$ (written without explicit argument $\vx$) be the
minimum value after convergence of the minimization procedure and
$\sx$ the value of the estimated parameters (the value of $\vx$ at the
minimum).  The covariance matrix of the estimates is then approximated
by the inverse of the curvature matrix at the minimum :
\begin{equation}
\Vx_{ij} = \left [\frac{1}{2}\frac{\partial^2 \chifullsix}
                {\partial x_i\partial x_j}\right ]^{-1}
\end{equation}

A traditional fit gives the track parameters $\sx$ {\em only} at the
{\em fixed} reference $z_0$, say at the beginning of the track.  In
order to find the best-fit track parameters at the {\em end} of the
track, one has to recompute $\Vd$ using the new reference and perform
a completely independent fit.  Due to multiple scattering, the results
of two fits using different reference $z_0$'s are {\em not} related by
the track model, and cannot be obtained from one another. This effect
is already present in a perfectly linear ideal case, where the
detector resolution does not depend on $\vx$. It just reflects the
fact that the weights of measurements to estimate the track parameters
depend (eventually strongly) on $z_0$.

This is unfortunate, since in practice, track extrapolation is often
desired from both ends of the track.  Furthermore, optimal track
estimates at every possible sensor position can be quite useful,
either to detect outliers efficiently, or to optimally collect hits
left over during a first pass.

Finally, if one attempts such a traditional track fit in an experiment
with large $N$ (such as NOMAD, where $N$ ranges up to 150) $N$ can be
too big to make inversion of $\Vd$ practical.

Given these difficulties, in NOMAD the Kalman filter was implemented
instead \cite{dcdoc}.

\section{The Kalman Filter}
\label{sec:kalman}
The Kalman filter is a least-squares stepwise parameter estimation
technique.  Originally developed in the early 60's to predict rocket
trajectories from a set of their past positions, it can be used to
handle multiple scattering while estimating track parameters.  We try
here to briefly shed light on the features of the Kalman filter for
track fitting and refer to the literature for more
details~\cite{Billoir84,Billoir85,Fruh,BilloirQian}.

The Kalman filter technique gives, mathematically speaking, exactly
the same result as a standard least squares minimization. In the
framework of track fitting, it essentially avoids big matrix inversion
and provides almost for free an optimal estimate of track parameters
at any location, allowing the detection of outlying measurements,
extrapolation and interpolation into other subdetectors.

The set of parameters $\vx$ is called the state vector in Kalman
filtering.

Starting from Eq.~\ref{eqn:sysb} we rewrite the system equation in a
stepwise form, where the state vector at location $k$ is obtained from
its value at previous location\footnote{One has to assume that the
measurements are ordered with respect to time to handle multiple
scattering because the covariance matrix of measurement residuals
depends on this order. Without multiple scattering, the ordering does
not affect the filter result as in the case of a traditional fit.}
$k-1$.
\begin{equation}
\vx_k \ = \ \vf_k(\vx_{k-1}) \ + \vw_k
\end{equation} 
     
We shall assume in the following that both $\vw_k$ and $\veps_k$ (the
measurement errors from Eq.~\ref{eqn:meas}) are independent random
variables with mean 0 and a finite covariance matrix.

Linearizing the system in the vicinity of $\vx_{k-1}$, one obtains:
\begin{equation}
 \vf_k(\vx_{k-1}) \ = \ F_k \cdot \vx_{k-1}
\end{equation}
\begin{equation}
 h_k(\vx_{k}) \ = \ H_k \cdot \vx_{k}
\end{equation}
for the track model and the measurement equation.

We can now recall keywords used in the Kalman filter estimation
technique:
\begin{itemize}
\item {\bf Prediction} is the estimation of the state vector at a 
``future" time, that is the estimation of the state vector at time or
position ($k+1$) using all the measurements up to and including $m_k$.
\item {\bf Filtering} is estimating the ``present" state vector 
based upon all present and ``past" measurements.  For {\em Forward}
filtering, this means estimating track parameters at $k$ using
measurements up to and including $m_k$. For {\em Backward} filtering,
this means estimating track parameters at $k$ using the measurements
$m_N$ down to $m_k$.
\item {\bf Filter.}
The algorithm which performs filtering is called a filter and is built
incrementally: filtering $m_1$ to $m_k$ consists in filtering $m_1$ to
$m_{k-1}$, propagating the track from $m_{k-1}$ to $m_k$ and including
$m_k$.  A filter can proceed forward ($k$ increases) or backward ($k$
decreases).
\item{\bf Smoothing} means using all the measurements to provide a track
parameter estimate at any position.  The smoothed estimate can be
obtained as a weighted mean of two filtered estimates: the first one
using $m_1$ to $m_k$ (forward), the other using $m_N$ to $m_{k+1}$
(backward).%
\footnote{This leads to a subtlety in practice, when we actually
have in hand the forward and backward filter estimates at $k$;
averaging these would lead to double-counting the information from
$m_k$.  Hence, to be proper, one must unfilter $m_k$ from one of the
estimates.  This small correction is implemented in our smoother and
in the quantity $\chikfb$ discussed below, but was deemed negligible
and never implemented in the other breakpoint quantities.}
\end{itemize}
  
One can understand the basic idea of the Kalman filter in the
following way.  If there is an estimate of the state vector at time
(location) $t_{k-1}$, it is extrapolated to time $t_k$ by means of the
system equation. The estimate at time $t_k$ is then computed as the
weighted mean of the predicted state vector and of the actual
measurement at time $t_k$, according to the measurement equation. The
information contained in this estimate can be passed back to all
previous estimates by means of a second filter running backwards or by
the smoother.

The main formulas for our linear dynamic system are the following:

{\bf System equation:}
\begin{equation}
 \vx_k \ = \ F_k \cdot \vx_{k-1} \ + \vw_k
\end{equation}
\begin{equation}
 E \{\vw_k \} \ = \ 0, \ \ \ \ {\rm cov} \{\vw_k \} \ = \ Q_k 
 \ \ \ \ (1 \leq k \leq N)
\end{equation}

{\bf Measurement equation:}
\begin{equation}
 m_k \ = \ H_k \cdot \vx_{k} \ + \ \veps_k
\end{equation}
\begin{equation}
 E \{ \veps_k \} \ = \ 0, \ \ \ \
{\rm cov} \{ \veps_k \} \ = \ V_k \ = \ G_k^{-1} 
\ \ \ \ (1 \leq k \leq N)
\end{equation}
where the matrices $Q_k$ and $V_k$ represent the process noise
(multiple scattering, bremsstrahlung, etc.) and measurement noise
(detector resolution) respectively. The details of $Q_k$ calculation
for the parameterization adopted in NOMAD can be found in
Ref.~\cite{Popov}. \\

As an example we include here the formulas for making a prediction:
\begin{itemize}
\item Extrapolation of the state vector: \\
$ \vx_{k}^{k-1} = F_k \vx_{k-1}$
\item Extrapolation of the covariance matrix: \\
$ C_k^{k-1} = F_k C_{k-1} F_k^T + Q_k$
\item Predicted residuals: \\
$ \vr_{k}^{k-1} = \vm_k - H_k \vx_k^{k-1}$
\item Covariance matrix of the predicted residuals: \\
$ R_{k}^{k-1} = V_k + H_k C_k^{k-1} H_k^T$
\end{itemize}

Using the Kalman filter, the computer time consumed for a track fit is
proportional to the number of hits on the track, while with the
traditional technique it is proportional to the cube of the same
number in case of multiple scattering.
 
After the Kalman fitting procedure one has the following available
information~:
\begin{itemize}
\item $\vxfmin$ and $\vxbmin$~: the Forward and Backward estimates 
of the state vector at position $k$, i.e., the estimate of the track
parameters at location $k$ using measurements 1 up to $k$ (forward) or
$N$ down to $k$ (backward).
\item 
$\chifmin$ and $\chibmin$~: the minimum $\chi^2$ value of the forward
and backward fits up to measurement $k$.
\item $\VxF$ and $\VxB$~: the covariance matrices of $\vxfmin$ 
and $\vxbmin$ respectively.
\item $\sx_{k}$, $\chifullsix$ and  $V^{(\sx_k)}$~:  the same 
quantities, determined from the smoothed estimates, the equivalent of
a full fit done at location $k$. Note that the $\chifullsix$ minimum
does not depends on the location at which the parameter are estimated
($\chifullsix$ for $\sx_{k}$ is independent of $k$).
\end{itemize}

Thus, much information exists as the by-product of a track fit: at
every hit on the track away from the ends, we have the results of
three fits for the track parameters at that hit: a fit to the part of
the track upstream, a fit to the part of the track downstream, and a
fit to the whole track.  This information is the input to breakpoint
analyses.

\section{Earlier Applications to Breakpoint Searches}
\label{sec:earlier}
A natural way to compare $\vxfmin$ and $\vxbmin$ is discussed by
R.~Fr\"uhwirth~\cite{Fruh} and was implemented in NOMAD tracking
\cite{dcdoc,Vo} before developing our extensions.  
One simply constructs the $\chi^2$ of the mismatch of all the
forward-backward parameters at each hit $k$:
\begin{equation}
\label{eqn:ck}
\chikfb = (\vxbmin - \vxfmin)^T \ [\VxB + \VxF]^{-1} 
 \ (\vxbmin - \vxfmin).
\end{equation}

The value of $\chikfb$ is easily computed from the following
relationship which holds for any $k$:
\begin{equation}
\label{eqn:chifullsixmin}
\chifullsix = \chifmin + \chibmin + \chikfb.
\end{equation}

After the track fit, we find the hit $k$ for which $\chikfb$ is a
maximum, i.e., for which the forward-backward mismatch in track
parameters has greatest significance; we call this maximum value
$\maxchikfb$.  One can assign a breakpoint at that $k$ there if
$\maxchikfb$ is above some threshold.  Fr\"uhwirth also investigated
various combinations of $\maxchikfb$ with $\chifmin$, $\chibmin$, and
degrees of freedom in the track fits, but concluded that $\maxchikfb$
was his best breakpoint test statistic \cite{Fruh,FK}.

\section{Some Additional Breakpoint Variables Based on Constrained Fits
to the Specific Breakpoint Types}
\label{sec:additional}

One may suspect that previously defined breakpoint tests do not make
optimal use of the available information.  Any $\chisq$ quantity is by
definition insensitive to the arithmetic sign of differences, while in
the processes of interest, a {\em decrease} in the momentum is
expected.  Furthermore, $\chikfb$ mixes all the parameter mismatch
information together.  The signed forward-backward mismatch in single
quantities such as $1/R$ can be examined, but it has the problem that
the other 5 parameters are not constrained to be the same.  (Physical
changes in $1/R$ can result, for example, in fitted mismatches in
$\phi$ as well as $1/R$.)  Finally, an optimal test should use a more
fully developed breakpoint hypothesis, so that a more meaningful
comparison of $\chisq$'s, with and without breakpoints, can take
place.

\subsection{Constrained Fits to Breakpoints}
\label{sec:constrained}
Here we show how to obtain and examine the result that one would get
by doing a traditional fit which uses {\em all} hits, {\em but which
allows a subset of the track parameters to be discontinuous} at a
particular hit $k$.  We parameterize the full track with 1 to 3 added
parameters in order to incorporate breakpoints of Types I, II, and III
at hit $k$.  E.g., for Type I, we replace the parameter $1/R$ by two
parameters, a forward value $1/R_F$ just before hit $k$ and a back
value $1/R_B$ just after hit $k$.  Thus, our fits to the 3 breakpoint
types have 7, 8, and 9 parameters, respectively.  We denote these sets
with breakpoints by $\valpha$ rather than $\vx$, and they are,
respectively for the three types:
\begin{eqnarray}
\valphaone &=& \{x,y,
                               1/R_F,1/R_B,
                               \tan\lambda,\phi,t\},
\label{eqn:alphaone}\\
\valphatwo &=& \{x,y,1/R,
                               \tan\lambda_F,\tan\lambda_B,
                               \phi_F,\phi_B,t\},
\label{eqn:alphatwo}\\
\valphathree &=& \{x,y,
                               1/R_F,1/R_B,
                               \tan\lambda_F,\tan\lambda_B,
                               \phi_F,\phi_B,t\}.
\label{eqn:alphathree}
\end{eqnarray}

For definiteness, we discuss here the concept in terms of a Type I
breakpoint. 

One can imagine a cumbersome procedure whereby one puts a Type I
breakpoint at a {\em particular} hit $k$, and performs a traditional
$\chisq$ track fit (with 7 parameters in our case) to {\em all} the
hits of the track, minimizing the full Type I track's $\chisq$, which
we call
\begin{equation}
\label{eqn:chifullsevenk}
 \chifullsevenk(\valphaone).
\end{equation}  
One would obtain the best estimate of $\valphaone$, its covariance
matrix, and the minimum value of $\chifullsevenk$, all for a
breakpoint at that particular hit $k$.  One could then repeat this for
each possible value of $k$, obtaining numerous potential
track-with-breakpoint fits.

Essentially {\em the same set of results} can be obtained far more
economically by starting from the results $\vxfmin$ and $\vxbmin$ at
each $k$ which {\em already exist} as by-products of the Kalman filter
fit.  These results carry all the information that we need, since
their error matrices contain the information (up to linear
approximation) on how $\chifmin$ and $\chibmin$ change when the track
parameters change.  We need only perform a linear $\chi^2$
minimization in which $\{\vxfmin,\vxbmin\}$ (now playing the role of
the ``measured data'') is compared to the Type I breakpoint model
prediction $H_I \valphaone$.  $H_I$ is the model matrix with 12 rows
and 7 columns containing only zeros and ones~:

\begin{equation}
H_I \valphaone  =  \{(x,y,1/R_F,\tan\lambda,\phi,t),
                     (x,y,1/R_B,\tan\lambda,\phi,t)\}.
\label{eqn:preddef}
\end{equation}
Or, introducing the two $6\times 7$ submatrices $H^F_I$ and $H^B_I$ of
$H_I$~:
\begin{equation}
H_I \valphaone =  (H^F_I \valphaone, H^B_I \valphaone).
\label{eqn:alphadef}
\end{equation}

Since our 12 pieces of ``measured data'' $\{\vxfmin,\vxbmin\}$ are two
independent sets of 6 parameters with their corresponding covariance
matrices, the appropriate chisquare can be written as~:
\begin{eqnarray}
\label{eqn:chifb}
\chikfb(\valpha) = 
    &(\vxfmin - H^F \valpha)^T \ [\VxF]^{-1} \ (\vxfmin - H^F \valpha) 
 \nonumber \\
  + &(\vxbmin - H^B \valpha)^T \ [\VxB]^{-1} \ (\vxbmin - H^B \valpha).
\end{eqnarray}
(Here and below, we suppress the subscript I since the equations are
true for all breakpoint types, with Eqns.~\ref{eqn:preddef} and
\ref{eqn:alphadef} suitably changed.)

The full $\chi^2$ of Eq.~\ref{eqn:chifullsevenk} can be written as~:
\begin{equation}
\chifullsevenk(\valpha) = (\mathbf{m} - 
\mathbf{h}\left (H \valpha \right))^T\left[ \Vd_k \right]^{-1}
(\mathbf{m} - \mathbf{h}\left( H \valpha \right ))
\label{eqn:fullchi2k}
\end{equation}
where $V^{(m)}_k$ is the block diagonal matrix containing the
covariance matrix $V^{(m,F)}_k$ of the measurements $(m_1\ldots m_k)$
and the covariance matrix $V^{(m,B)}_k$ of the measurement
$(m_{k+1}\ldots m_N)$.

As shown in Appendix~\ref{app-chisum}, $\chikfb(\valpha)$ of
Eqn.~\ref{eqn:chifb} is related to $\chifullsevenk(\valpha)$ of
Eqn.~\ref{eqn:fullchi2k} in a revealing way.  At each hit $k$,
\begin{equation}
\label{eqn:chifullseven}
\chifullsevenk(\valpha) = \chifmin + \chibmin + \chikfb(\valpha),
\end{equation}
subject to sufficient linearity in the fits.  Thus by finding the
minimum of $\chikfb(\valpha)$, we find the minimum of
$\chifullsevenk(\valpha)$, since $\chifmin$ and $\chibmin$ are known.

The minimum of $\chikfb(\valpha)$ is obtained {\em without iteration}
since the model relating $\valpha$ to $\{\vxfmin,\vxbmin\}$ is
linear. The set of estimated parameters is given by~:
\begin{equation}
\widehat{\valpha}_k = V^{(\widehat{\valpha}_k)}H^T 
                    ( V^{(\widehat{\vx})}_k )^{-1}\,{\bf X}
\end{equation}
where ${\bf X} = \{\vxfmin,\vxbmin\}$; $V^{(\widehat{\vx})}_k$ is the
block diagonal matrix containing $\VxF$ and $\VxB$; and where the
covariance matrix for the new estimate, $V^{(\widehat{\valpha}_k)}$,
is given by~:
\begin{equation}
V^{(\widehat{\valpha}_k)} = 
  \left [H^T (V^{(\widehat{\vx})}_k )^{-1}\, H \right ]^{-1}.
\end{equation} 
The value of $\chikfb$ at its minimum is
\begin{equation}
\chikfb(\widehat{\valpha}) = 
   - \left( H^T ( V^{(\widehat{\vx})}_k )^{-1}\,{\bf X}
\right)^T\cdot\widehat{\valpha}.
\end{equation}

Since the elements of $H$ are mostly 0, and the rest equal to 1, the
multiplications by $H$ and $H^T$ were done by hand before coding the
software; elements of $(V^{(\widehat{\vx})}_k )^{-1}$ have only one or
two simple terms.

The computer time to perform the Type I, II, and III breakpoint fits
at all hits $k$ (away from the track ends) was a negligible addition
(few per cent) to the NOMAD track finding and fitting software. This
added time was more than paid back by the speedup in matrix inversion
which was obtained by explicitly unrolling the loops in the DSINV
routine from CERNLIB~\cite{CERNLIB}.

\subsection{Breakpoint Variables Based on the Constrained Fits}
\label{sec:variables}
From the wealth of information thus available at each hit, we discuss
two of the most useful categories: 1) signed differences, in sigma, of
breakpoint parameters, and 2) $\chisq$ comparisons based on the Fisher
$F$-test.

As an example, for Type III breakpoint fits, we let $\DIIIR$ be the
forward-backward difference in $1/R$, divided by its standard
deviation (sigma, computed from the covariance matrix taking account
of errors in both quantities and their correlations).  This {\em
signed} quantity effectively gives the significance, in sigma, of the
jump in momentum at that hit.  Similar quantities, with analogous
notation, are calculated for all components of $\valpha$, for all
breakpoint Types.  Thus, for bremsstrahlung studies, $\DIR$ gives the
momentum change {\em under the constraint that all other track
parameters are continuous at hit $k$}.

The {\em Fisher $F$ statistic} \cite{Fisher2} is appropriate for
testing if adding parameters yields a statistically significant
reduction in the $\chisq$ of a fit.  It is simply the ratio of the
respective $\chisq$/dof for the two versions of the fits.  Thus we
naturally apply it to our track fits with and without breakpoint
parameters.  For example, for NOMAD's Type I fits, we have at each
hit,
\begin{equation}
\label{eqn:fisher7}
\FI = \big( \chifullsevenk/(N-7)\big) \ \Big/ \ 
          \big( \chifullsix/(N-6)\big);
\end{equation}
we similarly define $\FII$ and $\FIII$.

For a true breakpoint at a {\em given} hit $k$, each $F$ statistic in
principle obeys the standard significance-table distributions
\cite{Fisher2}.  However, we normally search through all hits in a
track for the lowest value of $F$, denoted $\minF$.  This lowest value
does not of course follow the usual significance-table values.  Hence,
as a practical matter, we use data or Monte Carlo events to measure
the distribution of $\minF$ and the effect of tests based on it.

\section{Effectiveness of the additional breakpoint variables}
\label{sec:results}

The effectiveness of any breakpoint search is of course highly
dependent on details of hardware (e.g., number and quality of the
position measurements) and software (e.g., how many tracks with
breakpoints are reconstructed as a single track).  We present for
illustration some experience with the NOMAD detector, using both
simulated and real data events.

\subsection{Electron identification and reconstruction}

\begin{figure}
\centering\epsfig{file=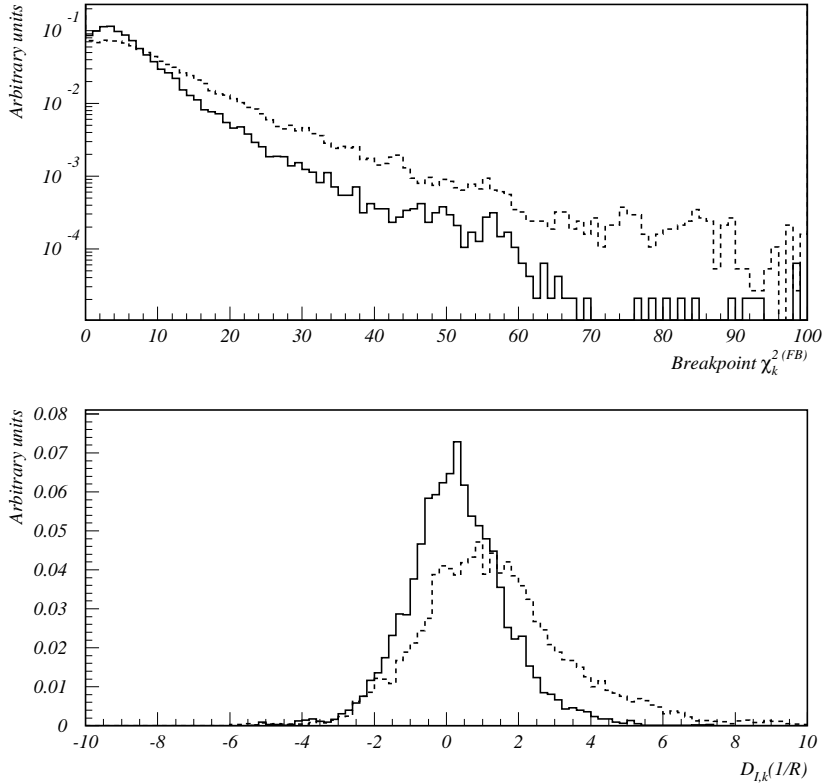,width=120mm}
\caption{
 Test of breakpoint search criteria using real data (muons producing
 $\delta$-electrons in the NOMAD setup).  Comparison of breakpoint
 chisquare ($\chikfb$) and normalized difference between curvatures in
 backward and forward directions ($(1/R_B - 1/R_F)/(\sigma_{1/R})$)
 for muons (solid line) and electrons (dashed line).  In both electron
 distributions, the excess on the right is evidence of potential
 breakpoints.}
\label{fig:ele_rd}
\end{figure}

Algorithms developed for electron identification and reconstruction
have been checked under running conditions using $\delta$-electrons
produced by straight-through muons (5 GeV $< p_\mu < 50$ GeV) crossing
the NOMAD detector during slow-extracted beam between neutrino spills.
This sample of selected electrons from real data can be used to check
the subdetector responses compared to simulations and to tune
breakpoint search criteria taking into account the effect of the drift
chambers alignment quality (see Fig.~\ref{fig:ele_rd}).

\begin{figure}
\vspace*{-6cm}
\epsfig{file=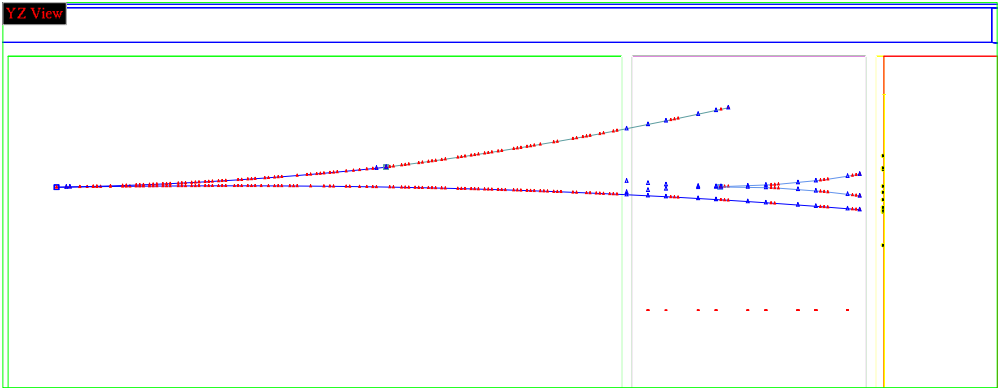,angle=-90,width=150mm}
\vspace*{-7cm}
\caption{
 A reconstructed event from real data (the projection onto the $yz$
 plane, in which tracks bend) before an attempt to apply the
 breakpoint search algorithm. The track at the bottom was identified
 as an electron by TRD. The triangles are track extrapolations used to
 search for more hits and to match information from different
 subdetectors.}
\label{fig:ELE_BREM_1}
\end{figure}
 
\begin{figure}
\vspace*{-6cm}
\epsfig{file=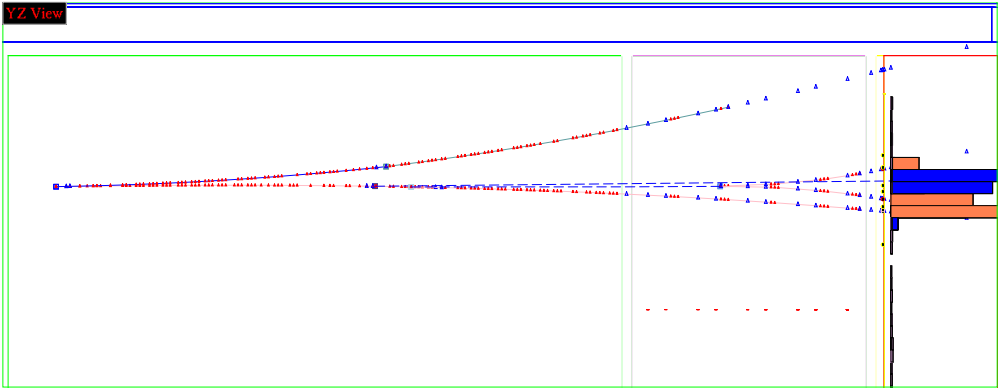,angle=-90,width=150mm}
\vspace*{-7cm}
\caption{The same event as before after applying a recursive 
 breakpoint search algorithm. Two breakpoints are found along the
 electron trajectory and they are associated with two photons (dashed
 lines): one built out of a conversion inside the drift chambers
 fiducial volume and the other from a stand-alone cluster in
 electromagnetic calorimeter.  The bars on the right are proportional
 to energy deposition in the electromagnetic calorimeter.  }
\label{fig:ELE_BREM_2}
\end{figure}

A special approach to deal with electrons emitting bremsstrahlung
photons (Type I breakpoints) in the NOMAD detector has been
developed~\cite{Popov}.  If one has identified a reconstructed track
in drift chambers as being an electron\footnote{Transition Radiation
Detector (TRD) is used for electron identification in the NOMAD
experiment.}, then hard bremsstrahlung photons can be looked for and
neutral tracks can be created requiring further matching with
preshower and electromagnetic calorimeter.  A successful application
of this approach to an event from real data is shown in
Fig.~\ref{fig:ELE_BREM_1} and Fig.~\ref{fig:ELE_BREM_2}.  A recursive
breakpoint search algorithm has been applied to a track identified as
an electron. As a result two breakpoints have been found, each
associated with an observed hard bremsstrahlung photon (one of which
converts to a reconstructed $e^+e^-$ pair).  When applying the
breakpoint search algorithm to an electron track one must keep in mind
a potential problem related to a possible presence of several kinks
along the trajectory (since it can bias the calculation of variables
used for the breakpoint search).  Details are in Ref.~\cite{Popov}.

\subsection{Studies using simulated pion decays}

As another example of using the breakpoint variables described in
Sec.~\ref{sec:variables}, we compare several tests for detecting the
Type III breakpoint (discontinuity of $1/R$, $\tan\lambda$, and
$\phi$) of the decay $\pimunu$, and for locating the position of the
decay.

The results shown here are for Monte Carlo (MC) simulation
\cite{NOMAD} of muon neutrino interactions in the NOMAD detector.
From the collaboration's standard MC samples, we selected two samples
of reconstructed tracks:
\begin{itemize}
\item 25000 pions which did not decay and which were reconstructed.
\item 2000 pions which did decay via $\pimunu$, and for which a 
{\em single} track was reconstructed consisting of hits left by {\em
both} the pion and the muon.
\end{itemize}
Neither of these two samples contains the pion decays which were
broken into separate tracks by the track-finding algorithm, with a
vertex assigned at the decay point.  The selected tracks were chosen
to have more than 20 hits ($N>20$) and no backward looping, and with
the MC pion decays within these hits.  Track-finding mistakes (for
example use of hits from other tracks in the event) were included, but
to obtain the pion track, we required that at least 90\% of the hits
be correctly assigned.

\begin{figure}
\centering\epsfig{file=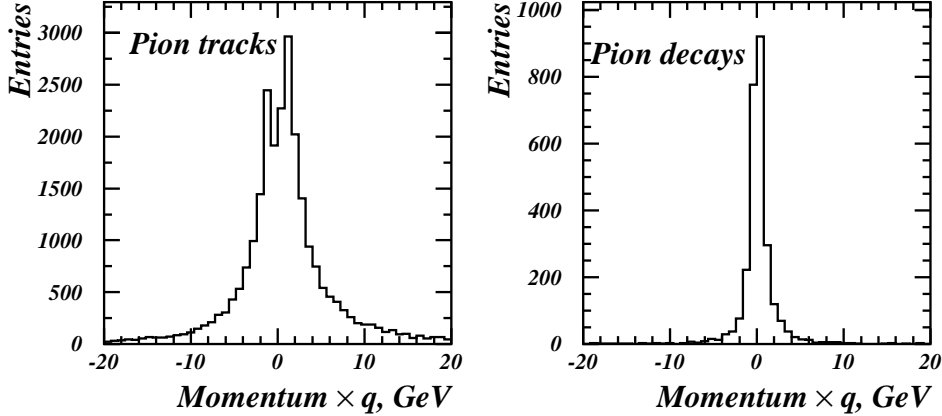,width=140mm}
\caption{Histogram of momentum$\times$charge for simulated pions 
without decay (left) and with decay (right).}
\label{fig:mom}
\end{figure}

Figure~\ref{fig:mom} contains histograms of momentum$\times$charge for
the two samples. In order to have samples with similar momenta,
we consider here only tracks with momentum less than 6 GeV.

\begin{figure}
\centering\epsfig{file=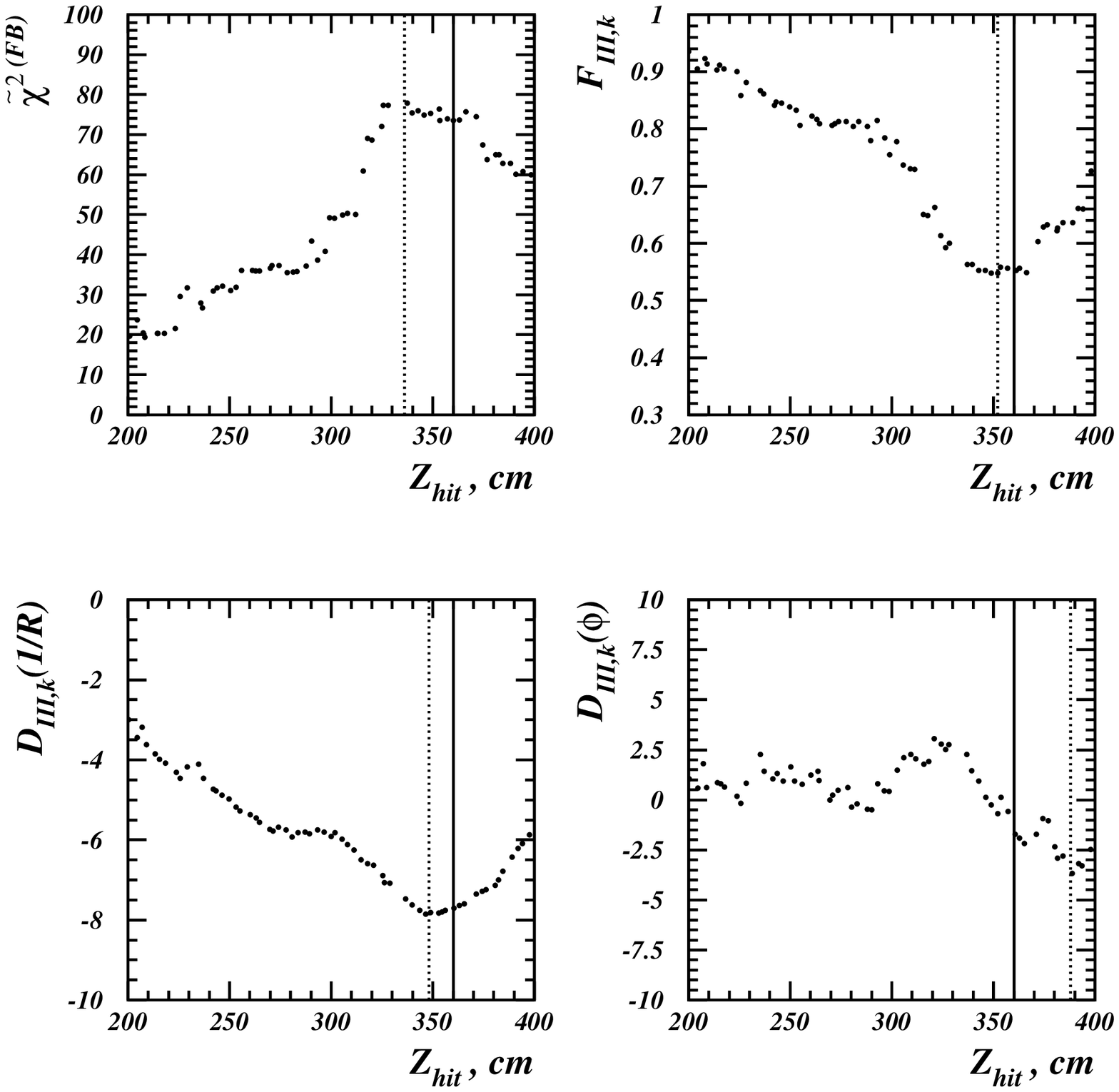,width=140mm}
\caption{$\maxchikfb$, $\FIII$, $\DIIIR$ and $\DIIIphi$ as a function 
of the $z$ of the DC hit. The MC decay point is shown in solid line.}
\label{fig:by_hit}
\end{figure}

Figure~\ref{fig:by_hit} shows, for one of the decaying negative pion
tracks, the values of $\chikfb$, $\FIII$, $\DIIIR$ and $\DIIIphi$ at
every hit where they are computed.  They are plotted at the
$z$-positions of the hits, which in the NOMAD detector are every few
cm.  The decay point in the MC is indicated by the line at 360 cm.
The dotted lines are at the $z$-positions of the extreme values
(maximum or minimum, as relevant for a breakpoint) of the respective
variables.  Near the MC decay point, both $\chikfb$ and $\FIII$ reach
their extreme values, indicating respectively: a large
forward-backward mismatch in the Kalman fits, and a marked improvement
in the track $\chisq$ by adding a three-parameter breakpoint.  The
sign of the change in $1/R$ corresponds to a decrease in momentum, as
expected in a decay. $\DIIIphi$ also shows an extremum near the 
breakpoint; however, we have generally not used it as the primary 
breakpoint indicator.

\subsection{Tests for Existence of a Breakpoint}
\label{sec:existence}
For testing the existence of a breakpoint, we compare the
following test statistics:
\begin{itemize}
\item $\chisq/\dof$ for the breakpointless fit.  (This test gives 
no additional information on the location of the breakpoint.)
\item $\maxchikfb$.  The breakpoint is located at the maximum 
value of the forward-backward mismatch chisquare ($\chikfb$) for
breakpointless fit among hits in the track;
\item $\minF$.  The breakpoint is located at the
minimum value Fisher $F$ among hits in the track (applicable
separately to the different breakpoint types if desired).
\item $\minFIII$ combined with the forward-backward mismatch 
in radius of the curvature $\DIIIR$ or angle $\DIIIphi$.  $\minFIII$
gives the breakpoint location and $\DIIIR$, $\DIIIphi$ are computed at
that location.
\end{itemize}

\begin{figure}
\centering\epsfig{file=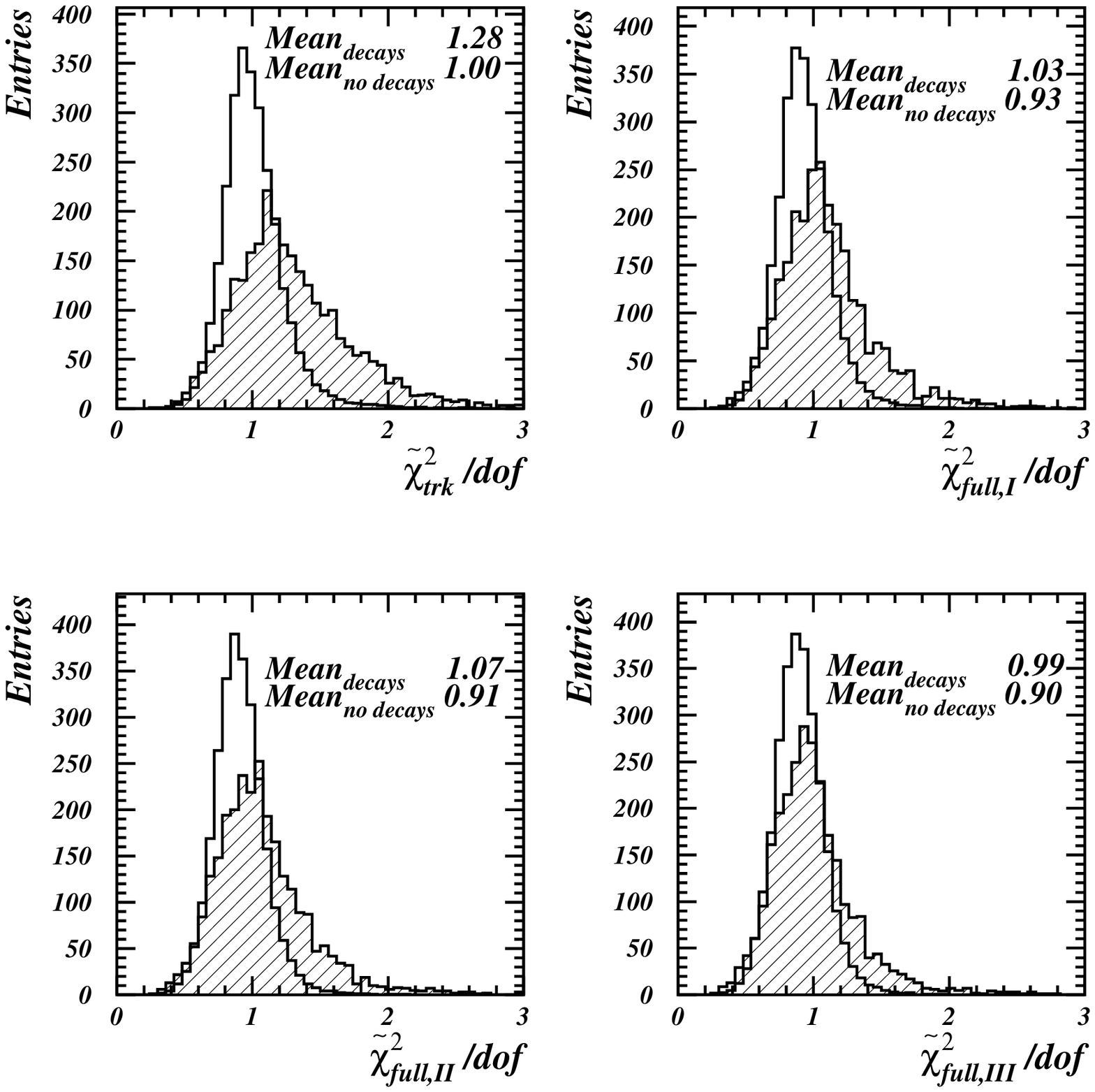,width=140mm}
\caption{Histograms of $\chifullsix/\dof$, $\minchifullseven/\dof$, 
$\minchifulleight/\dof$, and
$\minchifullnine/\dof$ for the sample of pion decays (shaded) and pions
with no decay (white).  The samples are normalized
to the same number of events.}
\label{fig:chi2_chi7_chi8_chi9}
\end{figure}

\begin{figure}
\centering\epsfig{file=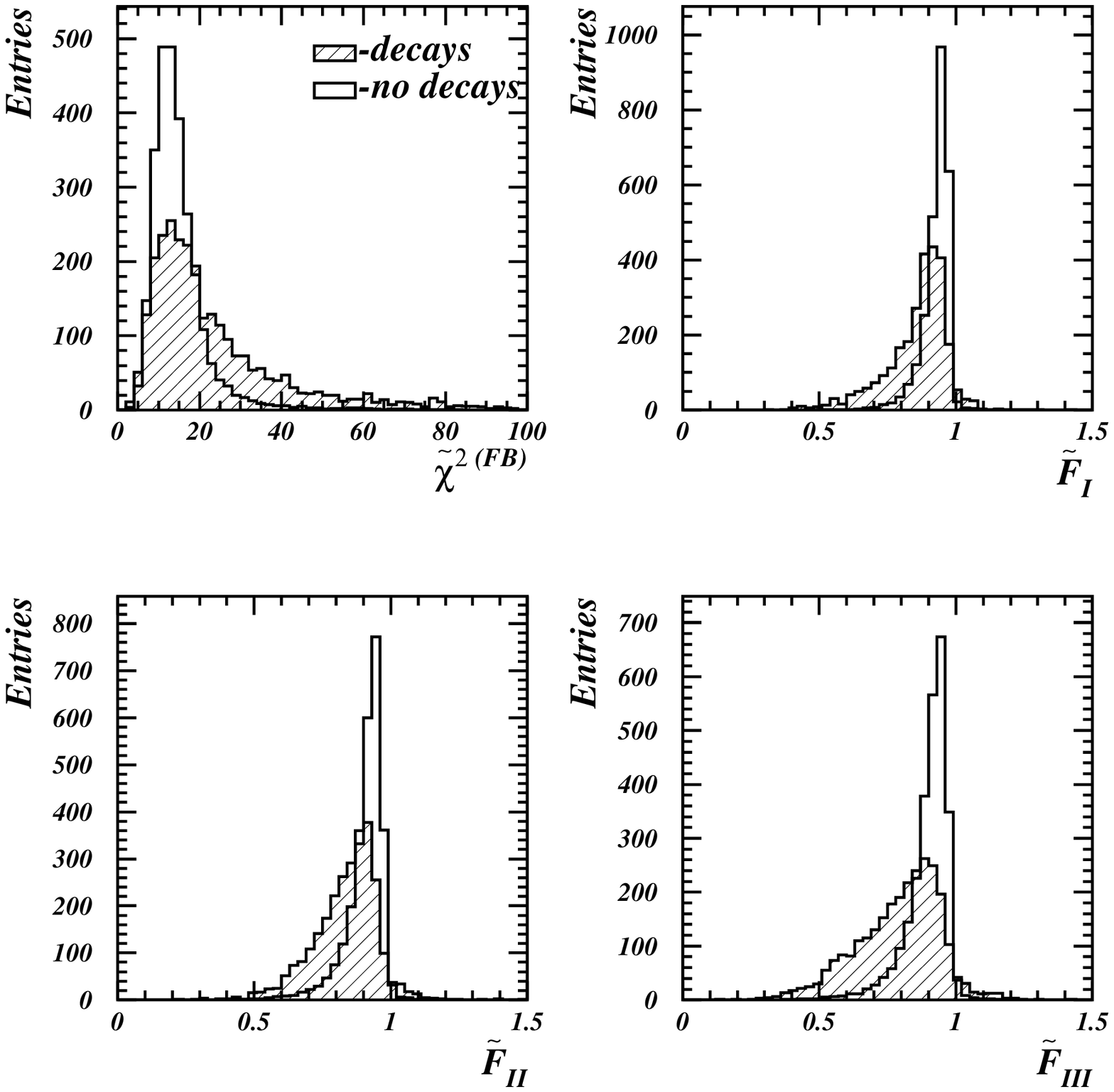,width=140mm}
\caption{Histograms of the $\maxchikfb$ and $\minFI$, $\minFII$, 
$\minFIII$ for pion decays (shaded) and pions with no decay (white).}
\label{fig:chi2_fchisq789}
\end{figure}

Fig.~\ref{fig:chi2_chi7_chi8_chi9} contains histograms of
$\chifullsix/\dof$, $\minchifullseven/\dof$, $\minchifulleight/\dof$,
and $\minchifullnine/\dof$ for decaying and non-decaying pions, with
the means shown.  The $\chisq/\dof$ is in general reduced by adding
breakpoint parameters.  A measure of the significance of the
improvement is the Fisher $F$ (Eq.~\ref{eqn:fisher7}), which is shown
in Fig.~\ref{fig:chi2_fchisq789} for the same fits.  Also shown in
Fig.~\ref{fig:chi2_fchisq789} is a histogram of $\maxchikfb$, the
maximum forward-backward mismatch from the breakpointless fit.

From the histograms in Fig.~\ref{fig:chi2_fchisq789}, one can
calculate the efficiency of labeling tracks as pion decays using
``cuts'' on these variables.  For tabulating a comparison, we choose a
cut value for $\minF$ such that 10\% of non-decaying pions are falsely
called pion decays.  We then find the efficiency for identifying real
pion decays, i.e., the percentage of pion decay tracks which are
called pion decays when using the same cut value.  This cut value in
principle depends on the dof, but for illustration we use the same cut
value for all track lengths.  The results are given in the first five
columns of Table~\ref{tab:chi2_fchisq789}.  We include for comparison
the result for testing for pion decay simply by using
$\chifullsix/\dof$, i.e., the test one would naturally use if one had
only a traditional non-Kalman track fit.  As observed by Fr\"uhwirth
\cite{Fruh}, this test is quite competitive with the $\maxchikfb$-test
for detecting the existence of a breakpoint, even though it gives no
information about the location.  The highest efficiency is obtained
using $\minFIII$.  We find this to be true for various other
comparisons, although one is cautioned that the particular
efficiencies listed are for the sample of decaying pions which were
not detected by the standard track finding/fitting, and are hence
expected to be highly experiment-dependent.

\begin{table}
\caption{Efficiency of the detection of pion decays, using the various
test statistics.  The first five columns are for simple cuts on the
respective variables.  The last column is for a cut on a 2D
likelihood-ratio test for $\minFIII$ vs $\DIIIR$ described in the
text. The cut value for each test is chosen so that 10\% of
non-decaying pions are wrongly called decays.  The efficiency is
computed with respect to the sample which contains only tracks which
were not detected as decaying by the original track-finding/fitting
algorithms.}
\label{tab:chi2_fchisq789}
\begin{center}
\begin{tabular}{cccccc}
  \hline
$\chifullsix/\dof$&  $\maxchikfb$   & $\minFI$ & $\minFII$ & 
$\minFIII$ & $\minFIII$ vs $\DIIIR$ \\ \hline
45\%  &  41\%  &  38\%  &  40\%  &  49\%  &  56\%  \\ \hline
\end{tabular}
\end{center}
\end{table}

\begin{figure}
\centering\epsfig{file=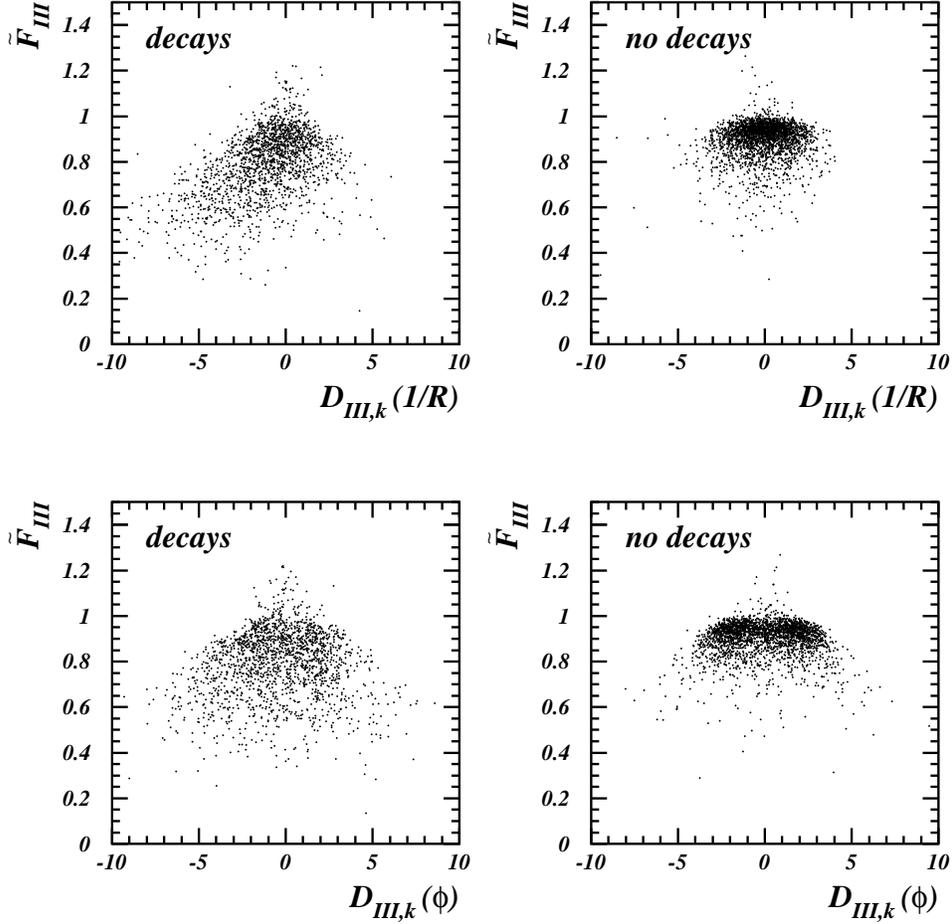,width=140mm}
\caption{$\minFIII$ vs $\DIIIR$ and $\minFIII$ vs $\DIIIphi$ 
histograms for the pion decays and pion
with no decays simulated tracks. In order to superimpose negative and
positive tracks on the same plots, we have changed the sign of
$\DIIIR$ for one charge. (Recall that 1/R is a signed quantity
reflecting the charge of the track.)  }
\label{fig:fchisq9_diff9rm1}
\end{figure}

Next we add the signed information available in the new fits: the
difference in radius of curvature $\DIIIR$, and/or the angular
differences $\DIIIphi$ and $\DIIIlam$.  We recommend that $\DIIIR$ not
be used to locate the breakpoint, but rather we evaluate this
difference at the location dictated by $\minFIII$.
Fig.~\ref{fig:fchisq9_diff9rm1} contains scatter plots of $\minFIII$
vs $\DIIIR$ for decays and non-decays.  Because true decays have a
decrease in momentum, a judicious cut on this scatter plot is more
effective than a cut solely on $\minFIII$.  An even more powerful
technique is to construct likelihood functions based on these 2D
densities, and use the constructed likelihood ratio as a test of pion
decay.  The result of such a procedure (applied with simple smoothing
of the 2D densities, and tested on an independent sample) is given in
the last column of Table~\ref{tab:chi2_fchisq789}.  The improvement
over previous breakpoint tests \cite{Fruh} is most significant.
Nearly as good efficiencies are obtained from scatter plots of
$\minFIII$ vs $\DIIIphi$ (also shown in 
Fig.~\ref{fig:fchisq9_diff9rm1}), and from $\minFIII$ vs $\DIIIlam$.

\subsection{Finding the Location of the Breakpoint}
\label{sec:location}
\begin{figure}
\centering\epsfig{file=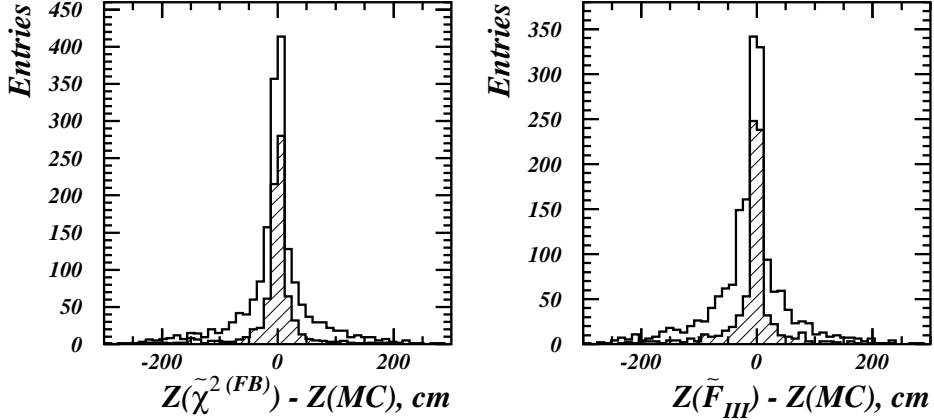,width=140mm}
\caption{$\maxchikfb$ and $\minFIII$ resolutions for MC 
pion decay sample in $z$. The white histograms are for the initial
sample; the shaded histograms are for events in which detected
breakpoints passed the selection criteria.}
\label{fig:res}
\end{figure}

One may also ask which of the variables gives the best determination
of the location of the breakpoint.  We studied in particular the
difference between the MC decay point and the $z$ position of the hit
corresponding to the extrema $\maxchikfb$ or $\minFIII$.  These are
histogrammed in Fig.~\ref{fig:res} for pions which decay.  The white
histograms show the resolution for the initial samples, with no
selection made using $\maxchikfb$ or ($\minFIII$ vs $\DIIIR$).  The
shaded histograms show the (more relevant) resolutions for tracks
remaining after decay selection using the respective variables.  There
was not a significant difference, within the limited scope of this
study.

\section{Conclusions}
\label{sec:conclusion}
Replacement of mismatch chisquare for all the forward-backward
parameters by the breakpoint variables introduced in
Sec.~\ref{sec:variables} can give added power to breakpoint detection
in the framework of Kalman filtering technique.  We show in particular
above that this is the case in a realistic simulation of pion decays
in the NOMAD detector.  In addition, these breakpoint variables have
been successfully used to reconstruct electron hard bremsstrahlung in
real data.  As expected on theoretical grounds, our most powerful
breakpoint detection is based on a scatter plot of a Fisher $F$-test
vs.\ an appropriate signed difference of a track parameter across the
breakpoint.

\begin{ack}
This work was performed within the NOMAD collaboration and hence
benefited from numerous aspects of NOMAD's simulation and event
reconstruction codes.  The authors are grateful to Emmanuel Gangler,
Kyan Schahmaneche, and Jean Gosset for their contributions to the
NOMAD Kalman filter.  We
gratefully acknowledge some early investigations by Mai Vo \cite{Vo}
on track breakpoints in NOMAD.
\end{ack}

\appendix
\section {\boldmath Derivation of Eq.~\protect\ref{eqn:chifullseven}:
$ \chifullsevenk = \chifmin + \chibmin + \chikfb$ }
\label{app-chisum}
In Eqn.~\ref{eqn:fullchi2k}, $V^{(m)}_k$ is the block diagonal matrix
containing the covariance matrix $V^{(m,F)}_k$ of the first $k$
measurements $\vddataF = (m_1\ldots m_k)$ and the covariance matrix
$V^{(m,B)}_k$ of the last $N-k$ measurements 
$\vddataB =(m_{k+1}\ldots m_N)$.  
The right-hand side of Eqn.~\ref{eqn:fullchi2k} can thus be split into
two terms:
\begin{eqnarray}
\chifullsevenk(\valpha) & = &\;\;\;
           \left [\vddataF - \mathbf{h}(H^F\valpha)\right ]^T\ 
[\VdF]^{-1}\left [\vddataF - \mathbf{h}(H^F\valpha)\right ] \nonumber \\
& & +  
           \left [\vddataB - \mathbf{h}(H^B\valpha)\right ]^T\ 
[\VdB]^{-1}\left [\vddataB - \mathbf{h}(H^B\valpha)\right ],
\label{eqn:chib}
\end{eqnarray}
where one recognizes, in analogy with Eqn.~\ref{eqn:chifull}, the
forward and backward $\chi^2$ terms.  We can expand each around their
respective minima $\vxfmin$ and $\vxbmin$, and recall that covariance
matrices are the inverse of curvature matrices, giving ~:
\begin{eqnarray}
\chifullsevenk(\valpha) & = &\;\;\;
\chifmin + (\Delta\vxf)^T \ [\VxF]^{-1} \ \Delta\vxf \nonumber \\
& & + \chibmin + (\Delta\vxb)^T \ [\VxB]^{-1} \ \Delta\vxb,
\label{eqn:chif-Taylor}
\end{eqnarray}
where $\Delta \vxf = \vxfmin - H^F \valpha$ and $\Delta \vxb = \vxbmin
- H^B \valpha$.\\ Combining this with Eqn.~\ref{eqn:chifb} yields
Eq.~\ref{eqn:chifullseven}.

\begin{thebibliography}{99}
\bibitem{FK}
Fr\"uhwirth refers to $\maxchikfb$ as $C_k$.  He defines a related
quantity, $F_k = ( C_k / n_1 ) / ( ( \chifmin + \chibmin ) / n_2 )$,
where $n_1$ is the d.o.f. for $\maxchikfb$ and $n_2$ is the sum of the
d.o.f. for $\chifmin$ and $\chibmin$.  Fr\"uhwirth concludes that the
$F_k$-test is less powerful than the $C_k$-test, and that the
$\chifullsixmin$ is almost as good an indicator of the existence of a
kink as the $C_k$-test.
\bibitem{CERNLIB} CERN Program Library routine number F012.
A modified version of the routine was used to write the source code
of the unrolled loops.
\end{thebibliography}
\end{document}